\documentclass[reprint,superscriptaddress,showpacs,amsmath,amssymb,aps,prl,floatfix,]{revtex4-1}
\usepackage{graphicx}
\usepackage{dcolumn}
\usepackage{color}
\usepackage{hyperref}
\usepackage{bm}
\graphicspath{{Images/}}
\begin{document}
	\preprint{APS/123-QED}
	\title{An ideal Weyl semimetal induced by magnetic exchange}
	\author{J.-R. Soh}
	\affiliation{Department of Physics, University of Oxford, Clarendon Laboratory, Parks Road, Oxford OX1 3PU, UK}%
	\author{F. de Juan}
	\affiliation{Department of Physics, University of Oxford, Clarendon Laboratory, Parks Road, Oxford OX1 3PU, UK}
	\affiliation{Donostia International Physics Center, 20018 Donostia-San Sebastian, Spain}
	\affiliation{IKERBASQUE, Basque Foundation for Science, Maria Diaz de Haro 3, 48013 Bilbao, Spain}
	\author{M. G. Vergniory}%
	\affiliation{Donostia International Physics Center, 20018 Donostia-San Sebastian, Spain}
	\affiliation{IKERBASQUE, Basque Foundation for Science, Maria Diaz de Haro 3, 48013 Bilbao, Spain}
	\author{N. B. M. Schr\"oter}%
	\affiliation{Paul Scherrer Institute, WSLA/202, 5232 Villigen PSI, Switzerland}
	\author{M. C. Rahn}%
	\affiliation{Department of Physics, University of Oxford, Clarendon Laboratory, Parks Road, Oxford OX1 3PU, UK}
	\altaffiliation[Present address: ]{MPA-CMMS, Los Alamos National Laboratory, Los Alamos, New Mexico 87545, USA}%
	\author{D.\,Y.\,Yan}
	\affiliation{Beijing National Laboratory  for Condensed Matter Physics, Institute of Physics, Chinese Academy of Sciences, Beijing 100190, China}
	\author{J. Jiang}
	\affiliation{Department of Physics, University of Oxford, Clarendon Laboratory, Parks Road, Oxford OX1 3PU, UK}
	\affiliation{School of Physical Science and Technology, ShanghaiTech University, Shanghai 201210, China}
	\affiliation{Advanced Light Source, Lawrence Berkeley National Laboratory, Berkeley, California 94720, USA}
	\author{M. Bristow}
	\affiliation{Department of Physics, University of Oxford, Clarendon Laboratory, Parks Road, Oxford OX1 3PU, UK}
	\author{P. Reiss}
	\affiliation{Department of Physics, University of Oxford, Clarendon Laboratory, Parks Road, Oxford OX1 3PU, UK}
	\author{J. N. Blandy}
	\affiliation{Department of Chemistry, University of Oxford, Inorganic Chemistry Laboratory, Oxford, OX1 3QR, UK}
	\author{Y. F. Guo}
	\affiliation{School of Physical Science and Technology, ShanghaiTech University, Shanghai 201210, China}
	\affiliation{CAS Center for Excellence in Superconducting Electronics (CENSE), Shanghai 200050, China}
	\author{Y. G. Shi}
	\affiliation{Beijing National Laboratory  for Condensed Matter Physics, Institute of Physics, Chinese Academy of Sciences, Beijing 100190, China}
	\author{T.\,K.\,Kim}
	\affiliation{Diamond Light Source, Harwell Campus, Didcot, OX11 0DE, UK}
	\author{A. McCollam}
	\affiliation{High Field Magnet Laboratory (HFML-EMFL), Radboud University, 6525 ED Nijmegen, Nijmegen, The Netherlands}
	\author{S. H. Simon}
	\affiliation{Department of Physics, University of Oxford, Clarendon Laboratory, Parks Road, Oxford OX1 3PU, UK}
	\author{Y. Chen}
	\affiliation{Department of Physics, University of Oxford, Clarendon Laboratory, Parks Road, Oxford OX1 3PU, UK}
	\affiliation{School of Physical Science and Technology, ShanghaiTech University, Shanghai 201210, China}
	\author{A. I. Coldea}
	\affiliation{Department of Physics, University of Oxford, Clarendon Laboratory, Parks Road, Oxford OX1 3PU, UK}
	\author{A. T. Boothroyd}
	\email{andrew.boothroyd@physics.ox.ac.uk}
	\affiliation{Department of Physics, University of Oxford, Clarendon Laboratory, Parks Road, Oxford OX1 3PU, UK}%
	\date{\today}
	
	\begin{abstract}
We report theoretical and experimental evidence that EuCd$_2$As$_2$ in magnetic fields greater than 1.6\,T applied along the $c$ axis is a Weyl semimetal with a single pair of Weyl nodes.  \textit{Ab initio} electronic structure calculations, verified at zero field by angle-resolved photoemission spectra, predict Weyl nodes with wavevectors ${\bf k} = (0,0,\pm 0.03)\times 2\pi/c$ at the Fermi level when the Eu spins are fully aligned along the $c$ axis. Shubnikov--de Haas oscillations measured in fields parallel to $c$ reveal a cyclotron effective mass of $m_{\rm c}^\ast = 0.08$\,$m_{\rm e}$ and a Fermi surface of extremal area $A_{\rm ext} = 0.24$\,nm$^{-2}$, corresponding to 0.1\% of the area of the Brillouin zone. The small values of  $m_{\rm c}^\ast$ and $A_{\rm ext}$ are consistent with quasiparticles near a Weyl node. The identification of EuCd$_2$As$_2$ as a model Weyl semimetal opens the door to fundamental tests of Weyl physics.
	\end{abstract}

	\maketitle
Weyl semimetals (WSMs) exhibit exceptional quantum electronic transport due to the presence of  topologically-protected band crossings called Weyl nodes~\cite{Armitage2018,Burkov2016}. The nodes come in pairs with opposite chirality, but their number and location in momentum space is otherwise material specific.
	
Weyl nodes are distinct from other topological features of electron band structures in severals respects, including (1) the bulk bands that cross at a Weyl node are non-degenerate, (2) the associated Weyl fermions have a definite chirality, and (3) the Weyl nodes are protected against perturbations that do not couple the nodes~\cite{Wan2011,Armitage2018,Burkov2016}. Moreover, the individual nodes within a pair act as a source and a sink of Berry curvature, a topological property of the electronic wavefunctions which relates directly to several anomalous transport phenomena~\cite{Armitage2018,Burkov2014}.

Weyl semimetal phases in crystals require either broken spatial inversion symmetry, or broken time-reversal symmetry (TRS), or both. There are a number of experimental realizations of the first type (with broken inversion symmetry only), especially in the TaAs structural family~\cite{Xu2015a,Lv2015,Yang2015,Xu2015b,Xu2015c}, but magnetic WSMs (with broken TRS) are still rare. The few known candidates are complicated by multiple pairs of Weyl nodes and/or by extra (non-topological) Fermi surface pockets which shroud the Weyl nodes~\cite{Wan2011,Suzuki2016,Liu2018,Wang2016,Nakatsuji2015,Sakai2018}. Magnetic WSMs are important for fundamental studies of Weyl fermions because it is possible for such systems to have only a single pair of Weyl nodes which, due to inversion symmetry, are guaranteed to be at the same energy and so have a vanishing density of states. By contrast, WSMs formed by breaking inversion symmetry (but with TRS) have a minimum of four nodes which are in general separated in energy.

\begin{figure}[t!]
	\includegraphics[width=0.5\textwidth]{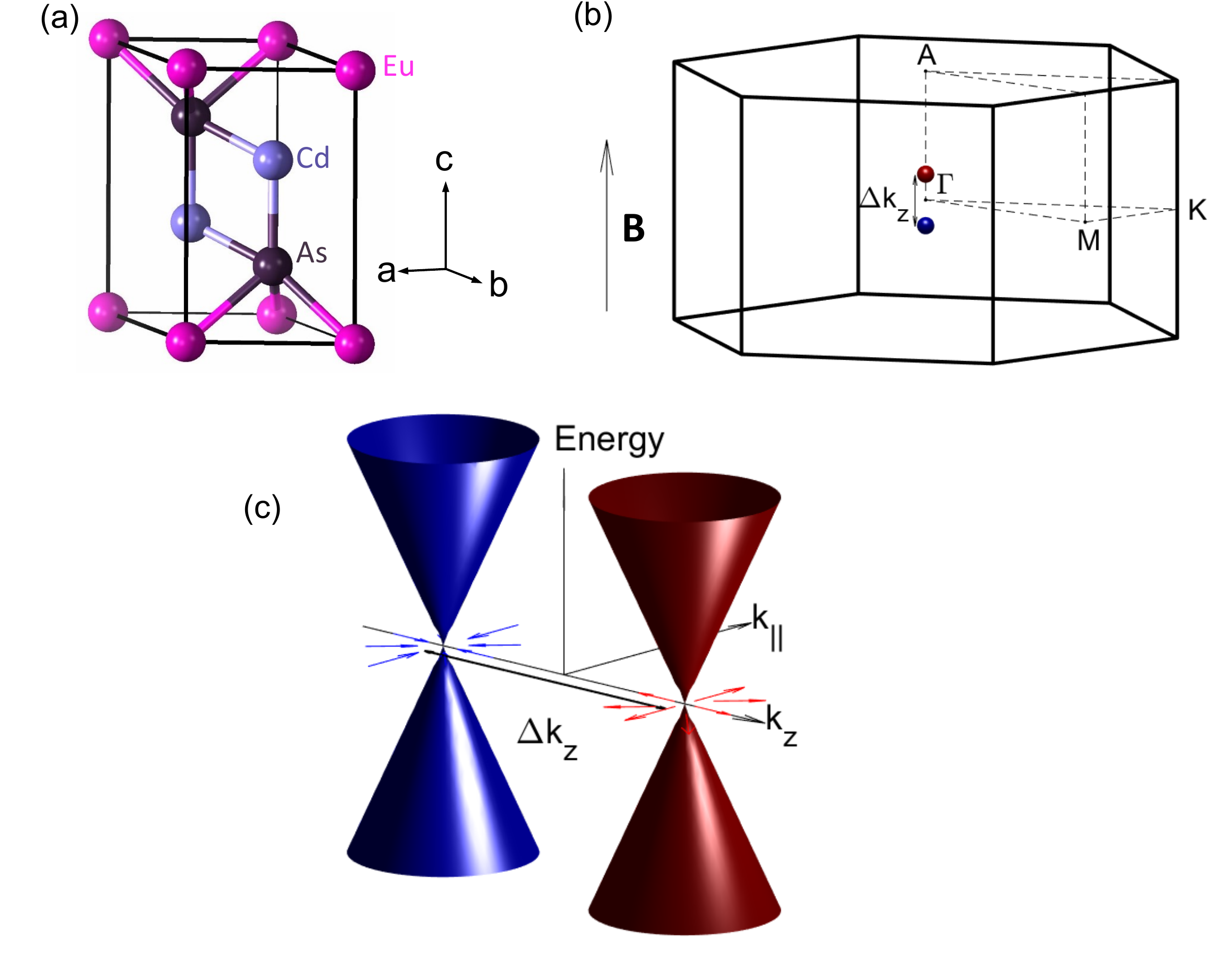}
	\caption{\label{fig:CK2_EuCd2As2_Figure_1} Crystal structure of EuCd$_2$As$_2$ and location of Weyl nodes in the Brillouin zone. (a) The trigonal unit cell for the space group $P\bar{3}m1$ (No.\,164). The Weyl fermions predominantly occupy orbitals in the double-corrugated Cd$_2$As$_2$ layers, which are sandwiched between the Eu layers~\cite{Rahn2018}. (b) The Weyl nodes lie along the A--$\Gamma$--A high symmetry line and are separated by $\Delta k_z$ (not shown to scale). (c) In the fully polarized state, singly degenerate conduction and valence bands meet at a pair of Weyl nodes (shown here schematically). The nodes act as a source and sink of Berry curvature (indicated by arrows).}
\end{figure}

 Following the initial discoveries~\cite{Xu2015a,Lv2015,Yang2015} there is now a need for better material realizations of WSMs, ideally comprising  a single pair of Weyl nodes located at or very close to the Fermi level and in an energy window free from other overlapping bands.  Here we propose the layered intermetallic EuCd$_2$As$_2$~\cite{WangHP2016,Rahn2018} to be such a system. We show that Weyl nodes in EuCd$_2$As$_2$ are magnetically-induced via exchange coupling, emerging when the Eu spins are aligned by a small external magnetic field applied along the $c$ axis.

		\begin{figure}[t!]
		\includegraphics[width=0.5\textwidth]{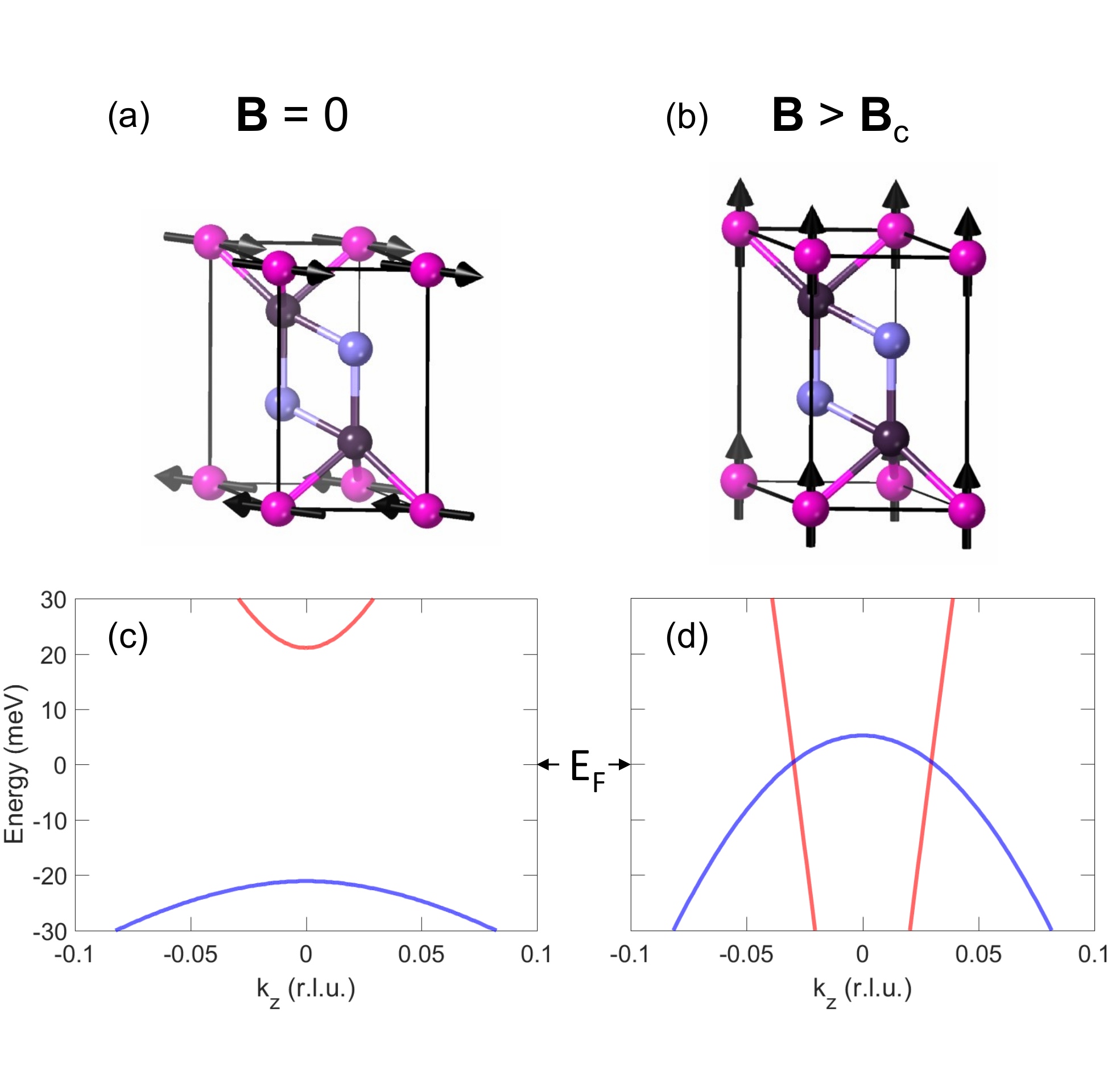}
		\caption{\label{fig:CK2_EuCd2As2_Figure_2} Exchange-induced Weyl nodes in EuCd$_2$As$_2$. (a), (c) In zero field, the Eu spins order in an A-type antiferromagnetic structure at $T < T_{\rm N}$, with the spins lying in the $ab$ plane. The corresponding band structure is gapped at $\Gamma$, and every band is doubly degenerate due to the combination of inversion and time-reversal symmetries. (b), (d) The Eu moments can be fully polarized along the $c$ axis in a small coercive field ($B>B_{\rm c}$), lifting the double degeneracy of the bands and creating a pair of Weyl nodes along $\Gamma$--A, at $k_z = 0.03$ r.l.u.}
	\end{figure}

Bulk single crystals of EuCd$_2$As$_2$ (ECA) were grown by a NaCl/KCl flux method~\cite{Schellenberg2011}. The  magnetic, transport and crystallographic properties of the crystals were fully consistent with earlier reports \cite{Rahn2018,WangHP2016,Schellenberg2011} (see Supplemental Material \cite{supp2019}). Magnetotransport measurements were performed on a Quantum Design Physical Properties Measurement System  for fields $B < 16$ T, and at the High Field Magnet Laboratory, Nijmegen, and the National High Magnetic Field Laboratory, Tallahassee, for fields up to 38\,T and 45\,T, respectively. ARPES measurements were performed at the high--resolution branch line of the beamline I05 at the Diamond Light Source, UK~\cite{Hoesch-2017}. 
A Scienta R4000 analyzer was used to select a photon energy of 130 eV, which approximately corresponds to the $k_z=0$ measurement plane.  The sample temperature was $T \sim 5$K. 

To calculate the band structure we employed density functional theory (DFT) as implemented in the Vienna \textit{Ab-initio} Simulation Package (VASP)~\cite{Kresse1996a,Kresse1996b}. The exchange correlation term is described according to the Perdew-Burke-Ernzerhof (PBE) prescription together with projected augmented-wave pseudopotentials~\cite{Perdew1996}. For the self-consistent calculations we used a 10$\times$10$\times$5  $k$-points mesh. The kinetic energy cut-off was set to 550\,eV. The spin-polarized band structures are calculated within GGA+U, with the value of U chosen to be 5\,eV to match the position of the Eu 4$f$ bands in the ARPES spectrum. The lattice parameters used in the calculations were $a = b = 0.443$\,nm and $c = 0.729$\,nm (see Supplemental Material~\cite{supp2019}).

The trigonal crystal structure of ECA, shown in Fig.~\ref{fig:CK2_EuCd2As2_Figure_1}(a), contains alternating layers of Eu$^{2+}$ and [Cd$_2$As$_2$]$^{2-}$ (Ref.~\cite{Artmann1996}).  The Eu ions carry a localized magnetic moment with spin $S = 7/2$ and essentially zero orbital angular momentum. Below $T_{\rm N} = 9.5$\,K the spins order in an A-type antiferromagnetic (AFM) structure in which the spins form ferromagnetic layers which stack antiferromagnetically along the $c$ axis [Fig.~\ref{fig:CK2_EuCd2As2_Figure_2}(a)] \cite{Rahn2018}. A relatively small magnetic field ($B_{\rm c} = 1.6$\,T at $T=2$\,K for $B \parallel c$) can be used to coerce the Eu spins into a fully aligned state~\cite{Rahn2018}. ECA is metallic at temperatures down to $T \sim 80$\,K, but for lower temperatures the resistivity increases to a sharp maximum at $T_{\rm N}$ before falling again at lower temperatures. This behaviour has been interpreted as due to scattering of conduction electrons by fluctuations of localized Eu magnetic moments which are exchange-coupled to the Cd and As orbitals~\cite{Rahn2018,WangHP2016}.

In previous \textit{ab initio} electronic structure calculations, where spins in the AFM state were found to be aligned with the $c$ axis, ECA was predicted to host a band inversion of doubly-degenerate As $4p$ and Cd $5s$ states near the Fermi level ($E_{\rm F}$), producing a crossing along the $\Gamma-A$ line protected by $C_3$ symmetry~\cite{Hua2018,Rahn2018}.  Experimentally~\cite{Rahn2018} (and in more recent calculations~\cite{Krishna2018}) the spins are in fact found to point perpendicular to the $c$ axis [Fig.~\ref{fig:CK2_EuCd2As2_Figure_2}(a)] breaking $C_3$ symmetry, so an avoided crossing at finite momentum would be expected.

We performed new \textit{ab initio} calculations of the electronic structure of ECA with the Eu spins configured (1) in the AFM state [Fig.~\ref{fig:CK2_EuCd2As2_Figure_2}(a)], and (2) in the ferromagnetic state with Eu spins fully aligned along the $c$ axis [Fig.~\ref{fig:CK2_EuCd2As2_Figure_2}(b)]. In the AFM state, the combination of time-reversal and inversion symmetry requires every band to be doubly degenerate. For a wide range of parameters, the AFM state displays a small direct gap at $\Gamma$ (Fig.~\ref{fig:CK2_EuCd2As2_Figure_2}(c) and Supplemental Material~\cite{supp2019}) which is mostly insensitive to the orientation of the spins.  When the Eu spins are fully spin-polarized along the $c$ axis the double degeneracy is lifted, and a single pair of Weyl nodes appears at $E_{\rm F}$ with no other Fermi surface pockets in the Brillouin zone  (Fig.~\ref{fig:CK2_EuCd2As2_Figure_2}(d) and Supplemental Material~\cite{supp2019}). These Weyl nodes lie along the $\Gamma - $A high symmetry line [see Fig.~\ref{fig:CK2_EuCd2As2_Figure_1}(b)] at wavevectors ${\bf k} = (0,0,\pm k_0)$ with $k_0 \simeq 0.03\times 2\pi/c = 0.26$\,nm$^{-1}$. ECA in a small magnetic field applied along the $c$ axis, therefore, is predicted to be a Weyl semimetal with a single pair of Weyl nodes located at $E_{\rm F}$.

The band splitting in the saturated phase is found to be $\sim$100\,meV, which is two orders of magnitude larger than the Zeeman splitting in the saturation field $B_{\rm c}$. From this we can conclude, first, that the calculations, which include exchange but no Zeeman interaction, are a good representation of the experimental situation in which a small magnetic field is used to align the Eu spins, and second, that the existence of the Weyl nodes is driven by exchange coupling to the Eu spins.

\begin{figure*}[t!]
	\includegraphics[width=0.8\textwidth]{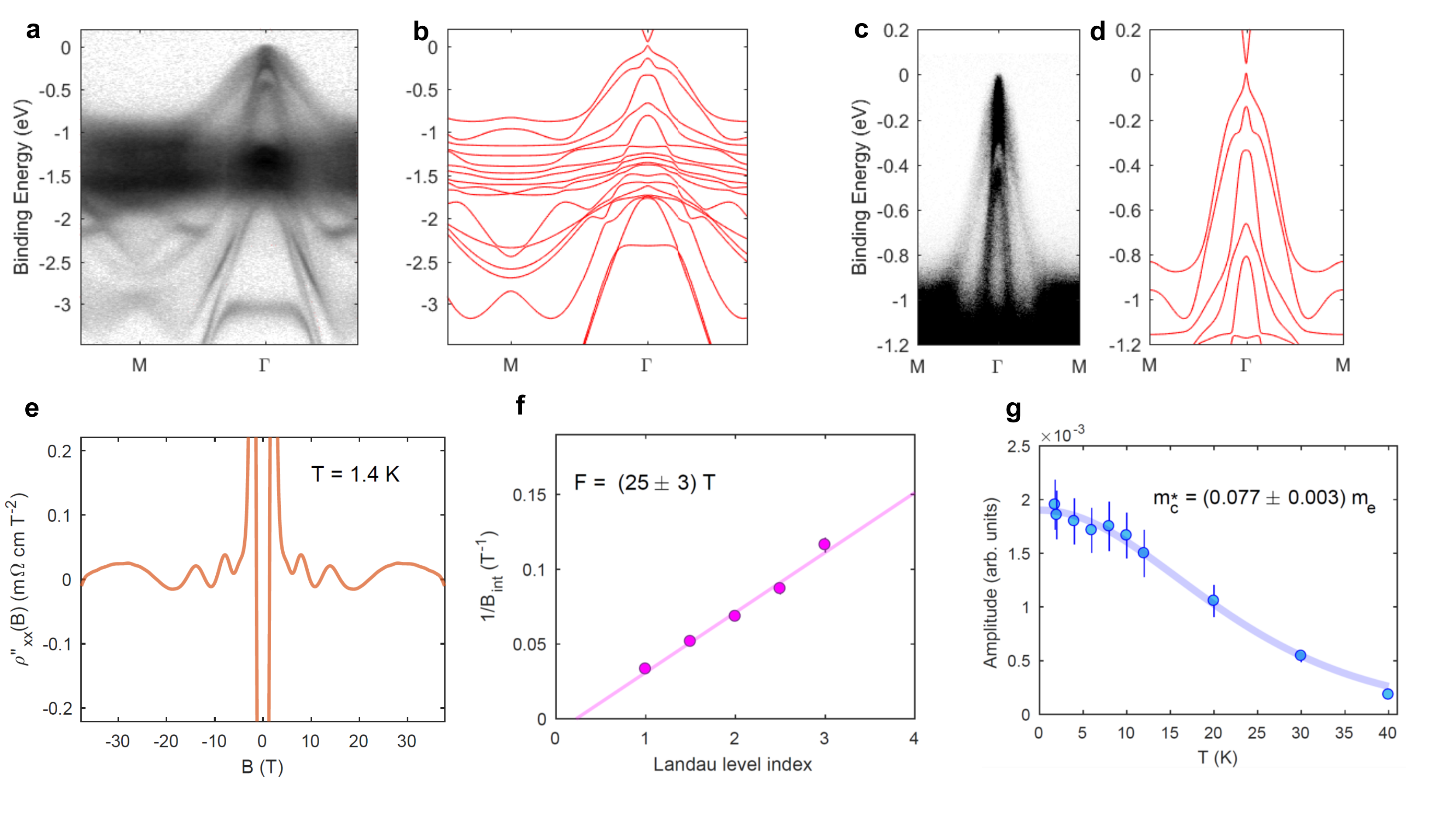}
	\caption{\label{fig:CK2_EuCd2As2_Figure_3} ARPES and high-field magnetotransport of EuCd$_2$As$_2$. (a), (c) ARPES spectrum as a function of wavevector along the M--$\Gamma$--M high symmetry line measured at $T\simeq 5$\,K with incident photon energy of 130\,eV. The data shown here are a sum of two measurements taken with linear-vertical and linear-horizontal polarization, respectively, to compensate the effect of selection-rules. Nonlinear scaling was applied to the intensity to enhance the visibility of bands with a small photoemission cross-section. (b), (d) Electronic bands calculated by DFT (in red). (e) The second derivative of the longitudinal resistivity $\rho''_{xx}(B)$ as a function of field applied along the $c$ axis. (f) Plot of $1/B_{\rm int}$ at the minima and maxima in $\rho_{xx}$ against Landau level index, with integers assigned to the minima, where $B_{\rm int} = B + \mu_0M$ is the internal field. The SdH frequency $F$ is obtained from the gradient of the linear fit shown.  (g) Temperature dependence of the amplitude in the SdH oscillation, measured at $B = 10$\,T. The line is a fit to the Lifshitz--Kosevich formula, from which the cyclotron effective mass $m_{\rm c}^\ast$ of the charge carriers is estimated. The quoted error in $m_{\rm c}^\ast$ derives from the least-squares fit, but the uncertainty in the measurement is expected to be larger because of the long period of the oscillations and the relatively narrow field range. }
\end{figure*}

In order to validate the \textit{ab initio} predictions we carried out angle-resolved photoemission spectroscopy (ARPES) and quantum oscillations measurements. ARPES data on ECA for $T < T_{\rm N}$ and zero applied magnetic field are presented in Figs.~\ref{fig:CK2_EuCd2As2_Figure_3}(a) and (c). These $k$--$E$ plots are for $k$ along the M--$\Gamma$--M path [see Fig.~\ref{fig:CK2_EuCd2As2_Figure_1}(b)] and show steeply dispersing bands approaching $E_{\rm F}$. The spectra are in good agreement with the \textit{ab initio} band structure [Figs.~\ref{fig:CK2_EuCd2As2_Figure_3}(b) and (d)] calculated for the observed AFM state with spins lying in the plane. The agreement is best when $E_{\rm F}$ is shifted slightly downward by about 50\,meV, which indicates that the sample is very slightly hole-doped. 

\begin{figure}[b]
	\includegraphics[width=0.5\textwidth]{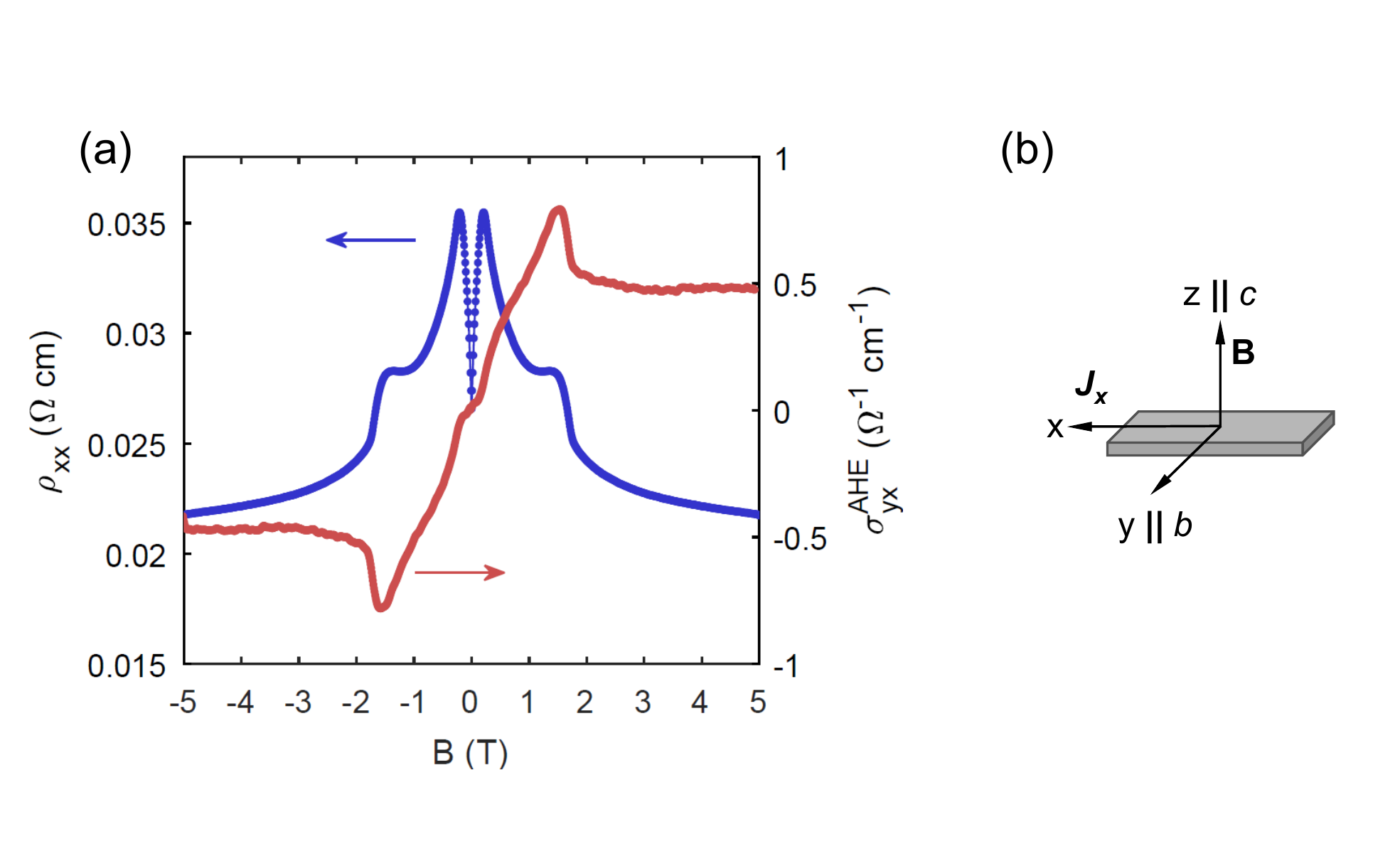}
	\caption{\label{fig:CK2_EuCd2As2_Figure_4} Magnetotransport of EuCd$_2$As$_2$. (a) $\rho_{xx}$ and $\sigma_{yx}$ as a function of field. (b) Definition of the $xyz$ axes relative to the crystal orientation and directions of the current and applied field used in the measurement.}
\end{figure}

Our quantum oscillations measurements are summarized in Figs.~\ref{fig:CK2_EuCd2As2_Figure_3}(e)--(g). Figure~\ref{fig:CK2_EuCd2As2_Figure_3}(e) shows the second derivative of the in-plane longitudinal resistance measured at $T = 1.4$\,K in magnetic fields applied parallel to the $c$ axis ($B \parallel c$) up to 37\,T, well above the coercive field (see Supplemental Material~\cite{supp2019} for details of the data treatment). For $B < B_{\rm c}$ there is a very strong variation in magnetoresistance associated with the progressive canting of the spins towards the $c$ axis, shown also in Fig.~\ref{fig:CK2_EuCd2As2_Figure_4}(a),  but at higher fields the curve displays clear Shubnikov--de Haas (SdH) oscillations. Only a single SdH oscillation frequency could be resolved, consistent with a single band. Moreover, we do not find any evidence for a spin-splitting of the Landau levels, in contrast to the SdH data on the structurally-related Dirac semimetal Cd$_3$As$_2$ (Ref.~\cite{Narayanan2015}). A lack of spin splitting would be consistent with the prediction that for $B>B_c$ the bands are already split by a constant exchange field, implying that the observed SdH oscillations correspond to the small pockets derived from the Weyl points when $E_{\rm F}$ is shifted downwards slightly, as suggested by the ARPES measurements.


The maxima and minima of the oscillations are plotted on a Landau level index plot in Fig.~\ref{fig:CK2_EuCd2As2_Figure_3}(f), with minima in $\rho_{xx}$ assigned to the integers~\cite{Ando2013}. The SdH frequency  obtained from the gradient is  $F = 25 \pm 3$\,T,  
which  converts via the Onsager relation $F = (\hbar/2\pi e)A_{\rm ext}$ to an extremal area of the Fermi surface perpendicular to the $c$ axis of $A_{\rm ext} = 0.24$\,nm$^{-2}$, or $k_{\rm F} = 0.28$\,nm$^{-1}$ assuming a circular cross-section.  This $A_{\rm ext}$ value represents approximately 0.1\% of the cross-sectional area of the Brillouin zone perpendicular to the $c$ axis.
Measurements at higher fields up to 45\,T did not find any additional oscillation frequencies (see Supplemental Material~\cite{supp2019}), and confirmed that the maximum centred on 30\,T in Fig.~\ref{fig:CK2_EuCd2As2_Figure_3}(d) corresponds to the quantum limit (first Landau level). 

Figure~\ref{fig:CK2_EuCd2As2_Figure_3}(f) shows the temperature dependence of the SdH oscillation amplitude up to 40\,K. By fitting the data to the Lifshitz--Kosevich formula (amplitude $\sim X/\sinh X$, where $X = 2\pi^2k_{\rm B}Tm_{\rm c}^\ast/e\hbar B$ and $m_{\rm c}^\ast = (\hbar^2/2\pi)\, {\rm d}A_{\rm ext}/{\rm d}E$) we obtain a cyclotron effective mass of about $m_{\rm c}^\ast = 0.08m_{\rm e}$.  The observation that $m_{\rm c}^\ast/m_{\rm e} \ll 1$ is consistent with quasiparticles near a Dirac or Weyl node. 
A small effective mass is also found in our \textit{ab initio} calculations. Assuming $k_{\rm F} = 0.28$\,nm$^{-1}$, as determined from the SdH data, and a circular Fermi surface cross-section, we obtain $m_{\rm c}^\ast = 0.18m_{\rm e}$ from the calculated band structure. We caution, however, that the measured and calculated $m_{\rm c}^\ast$ are not directly comparable because the Lifshitz--Kosevich formula assumes that the Landau levels are equally-spaced and that many levels are filled, neither of which applies here. 

Our quantum oscillations and ARPES results, as well as previous optical reflectivity measurements which found clear evidence for a very low carrier density~\cite{WangHP2016}, all point to a very small Fermi surface, and support the prediction that in the spin-polarized state of ECA there is a single pair of Weyl nodes located close to $\Gamma$ along $\Gamma - $A, in a small window of energy free from other bands. The small effective mass and Fermi surface area from the SdH data, together with $p$-type Hall transport~\cite{Rahn2018}, indicate that the crystals used in this study are slightly hole-doped. From the SdH measurements and \textit{ab initio} in-plane dispersion we estimate that $E_{\rm F}$ is located approximately 52\,meV below the Weyl node (see Supplemental Material~\cite{supp2019}), which is consistent with the shift applied to the DFT bands in order to match the ARPES data.

In recent years, there has been a great deal of interest in measuring anomalous transport effects caused by Berry curvature in topological semimetals, especially the anomalous Hall effect (AHE)~\cite{Nagaosa2010,Burkov2014}. In a Weyl semimetal, the Berry curvature is associated with the separation $\Delta k$ of the Weyl nodes in \textbf{k}-space, as illustrated in Fig.~1c, and for a single pair of nodes at $E_{\rm F}$ the anomalous Hall conductivity has the universal form~\cite{Burkov2014},
\begin{equation}
	\sigma_{yx}^{\rm AHE} = \frac{e^2}{2\pi h}\Delta k.
	\label{eq1}
\end{equation}
In experiments, $\Delta k$ is typically field-dependent due to the effect of field on the band splitting. This makes it difficult to separate the anomalous and semi-classical contributions to the Hall effect, as the latter is also field-dependent. In ECA, however, $\Delta k$ is almost constant for fields above the saturation field $B_{\rm c} =1.6$\,T, which makes it straightforward to isolate the anomalous part of the Hall resistivity. In principle, therefore, ECA is an ideal system for studying the AHE experimentally.

Figure~\ref{fig:CK2_EuCd2As2_Figure_4}(a) presents measurements of the longitudinal magnetoresistance $\rho_{xx}$ and the anomalous part of the transverse (Hall) conductivity $\sigma_{yx}^{\rm AHE}$ at $T = 2$\,K as a function of field applied parallel to the $c$ axis (the experimental geometry is shown in Fig.~\ref{fig:CK2_EuCd2As2_Figure_4}(b), and the procedure to obtain the anomalous part of $\sigma_{yx}$ is described in the Supplemental Material~\cite{supp2019}). There are rapid changes in $\rho_{xx}$ at low field due to the reorientation of the Eu spins in the magnetic field, as noted earlier, but above the saturation field $B_{\rm c} = 1.6$\,T $\rho_{xx}$ decreases monotonically with field. The field range in Fig.~\ref{fig:CK2_EuCd2As2_Figure_4}(a) is below that where quantum oscillations become observable [see Fig.~\ref{fig:CK2_EuCd2As2_Figure_3}(e)]. The $\sigma_{yx}^{\rm AHE}(B)$ curve is an odd function of field, increasing rapidly for $0 < B < B_{\rm c}$ and remaining constant for $B > B_{\rm c}$, consistent with a non-zero anomalous Hall conductivity.

Assuming $\Delta k \simeq 0.52$\,nm$^{-1}$ from our \textit{ab initio} results, equation~(\ref{eq1}) predicts the anomalous Hall conductance for ECA  to be $\sigma_{yx}^{\rm AHE} \simeq 30$\,$\Omega^{-1}$cm$^{-1}$, which is significantly larger than observed. This prediction, however, applies only when the Weyl nodes lie exactly at $E_{\rm F}$. In the samples used here the nodes are slightly shifted from $E_{\rm F}$, and in this situation other factors are expected to affect the Berry phase~\cite{Nagaosa2010}.  One such factor is disorder. We have found the AHE in ECA to be reduced by the polishing process used to shape the Hall bar samples. Although it has been argued that disorder-induced contributions to the AHE are absent when $E_{\rm F}$ is near the nodes~\cite{Burkov2014}, the presence of a tilt in the dispersion makes these contributions possible in the form of skew scattering~\cite{Steiner2017,Mukherjee2018}. The significant tilt predicted in our \textit{ab initio} calculations [Fig.~\ref{fig:CK2_EuCd2As2_Figure_2}(d)] might explain why the AHE is so reduced.

The simple structure of the Weyl nodes in ECA makes it an ideal material with which to study the different contributions to the AHE. This could be achieved by tuning the position of $E_{\rm F}$ relative to the Weyl nodes by doping or application of pressure, or by controlling the level of defects by irradiation. The degrading effects of polishing could be avoided by studying transport phenomena with thin film samples.

More generally, ECA could provide the means to test predictions of other exotic physics in Weyl semimetals, such as the anomalous Nernst and thermal Hall effects~\cite{Sakai2018,Ferreiros2017}, 
non-reciprocal effects in light propagation~\cite{Kotov}, the repulsive Casimir effect \cite{Wilson}, or to probe the effects of the chiral anomaly in the optical absorption~\cite{Ashby2014} and non-local transport~\cite{Sid2014}.

\textit{Note added:} Very recently, Ma \textit{et al.}~\cite{Ma2019} reported ARPES results on EuCd$_2$As$_2$ which provide evidence for the existence of dynamically fluctuating Weyl points due to ferromagnetic correlations present in zero field at $T > T_{\rm N}$. The results in Ref.~\onlinecite{Ma2019} provide experimental confirmation that the DFT calculations correctly predict the effects of ferromagnetic polarization of the Eu spins on the band structure.

\begin{acknowledgements}
The work at Oxford was supported by the U.K. Engineering and Physical  Sciences  Research  Council (EPSRC)  (Grant Nos. EP/I004475/1, EP/N034872/1, EP/M020517/1 and EP/N01930X/1), the John Fell Fund, and the Oxford Centre for Applied Superconductivity.  J.-R.S. acknowledges support from the Singapore National Science Scholarship, Agency for Science Technology and Research. F.d.J. acknowledges funding from the European Union's Horizon 2020 research and innovation programme under the Marie Sklodowska Curie grant agreement No. 705968. 
M.C.R. was supported by the Oxford University Clarendon
Fund, the Los Alamos National Laboratory Director’s Fund, and the Alexander von Humboldt-Stiftung.  Crystal growth was carried out with support from the National  Key Research  and  Development  Program of China (grant nos. 2017YFA0302901 and 2016YFA0300604) and the Strategic Priority Research Program  (B) of  the  Chinese  Academy  of  Sciences  (Grant No. XDB07020100).  We acknowledge the Diamond Light Source for time on beamlines I11 (proposal EE18786) and I05 (proposal SI19234).
Work at the HFML, a member of the European Magnetic Field Laboratory (EMFL), was supported in part by the EPSRC via its membership of the EMFL (grant no. EP/N01085X/1). Work at the NHMFL was sponsored by the National Science Foundation under Cooperative Agreement No. DMR-1157490, and by the State of Florida. We thank D. Graph for technical support at the NHMFL.
\end{acknowledgements}


\nocite{*}
\bibliography{library}
\end{document}